\documentclass[aps,showpacs,amsmath,amssymbd,dvips,showkeys,preprintnumbers]{revtex4}
\usepackage{graphicx}   
\usepackage{color}      
\usepackage{braket}
\usepackage{tikz}
\usepackage{pgfplots}
\def \be  {\begin{equation}}
\def \ee  {\end{equation}}
\def \ee  {\end{equation}}
\def \bea {\begin{eqnarray}}
\def \eea {\end{eqnarray}}

\begin{document}

\preprint{ECTP-2019-09}    
\preprint{WLCAPP-2019-09}
\hspace{0.05cm}
\title{Deconfinement and freezeout boundaries in equilibrium thermal models} 

\author{Abdel Nasser Tawfik} 
\email{atawfik@nu.edu.eg}
\affiliation{Nile University, Egyptian Center for Theoretical Physics (ECTP), Juhayna Square of 26th-July-Corridor, 12588 Giza, Egypt}
\affiliation{Goethe University, Institute for Theoretical Physics (ITP), Max-von-Laue-Str. 1, D-60438 Frankfurt am Main, Germany}

\author{Muhammad Maher} 
\affiliation{Helwan University, Faculty of Science, Physics Department, 11795 Ain Helwan, Egypt}

\author{A. H. El-Kateb} 
\affiliation{Helwan University, Faculty of Science, Physics Department, 11795 Ain Helwan, Egypt}

\author{Sara Abdelaziz} 
\affiliation{Helwan University, Faculty of Science, Physics Department, 11795 Ain Helwan, Egypt}

\date{\today}

\begin{abstract}

In different approaches, the temperature-baryon density plane of QCD matter is studied for deconfinement and chemical freezeout boundaries. Results from various heavy-ion experiments are compared with the recent lattice simulations, the effective QCD-like Polyakov linear-sigma model, and the equilibrium thermal models. Along the entire freezeout boundary, there is an excellent agreement between the thermal model calculations and the experiments. Also, the thermal model calculations agree well with the estimations deduced from the Polyakov linear-sigma model (PLSM). At low baryonic density or high energies, both deconfinement and chemical freezeout boundaries are likely coincident and therefore the agreement with the lattice simulations becomes excellent as well, while at large baryonic density, the two boundaries become distinguishable forming a phase where hadrons and quark-gluon plasma likely coexist.

\end{abstract}

 
\maketitle

\section{Introduction}
\label{sec:Intr}

Strongly interacting matter under extreme conditions is characterized by different phases and different types of the phase transitions \cite{Banks:1983me}. The hadronic phase, where stable baryons build up a great part of the Universe and the entire everyday life, is a well known phase. At high temperatures and/or densities, other phases appear. For instance, at temperatures of a few MeV, chiral symmetry restoration and deconfinement transition take place, where quarks and gluons are conjectured to move almost freely within colored phase known as the quark-gluon plasma (QGP)  \cite{Tawfik:2014eba}. At low temperatures but large densities, the hadronic (baryonic) matter forming compact interstellar objects such as neutron stars is indubitably observed in a conventional way and very recently gravitational waves from neutron star mergers have been detected, as well \cite{TheLIGOScientific:2017qsa}. At higher densities, extreme interstellar objects such as quark stars are also speculated \cite{Ivanenko1965}. In lattice quantum chromodynamics (QCD), different orders of chiral and deconfinement transitions have be characterized, especially at low baryon densities. 

The program of heavy-ion collision experiments dates back to early 1980's. Past (AGS, SIS, SPS), current (RHIC, LHC), and future facilies (FAIR, NICA) help in answering essential questions about  the thermodynamics of the strongly interacting matter and in mapping out the temperature-baryon density plane \cite{Tawfik:2014eba}. The unambiguous evidence on the formation of QGP is an example of a great imperical achievement \cite{Gyulassy:2004zy,Busza:2018rrf}. The colliding nuclei are conjectured to form a fireball that cools down by rapid expansion and finally hadronizes into individual uncorrelated hadrons. The present script focuses on the temperature-baryon density plane, concretely near the hadron-QGP boundaries, in framework of equilibrium thermal model \cite{Greiner:2004vm}. To this end, we put forward a basic assumption that both directions, hadron-QGP and QGP-hadron phase are quantum-mechanically allowed \cite{Muller:2017vnp}. In other words, the picture drawn so far seems in fundamental conflict with the time arrow. The concept of arrow of time prevents the reverse direction, especially if the change in the degrees of freedom or entropies aren't following the causality principle; second law of thermodynamics. The statistical thermal approaches work well near to both deconfinement and chemical freezeout boundaries \cite{Vovchenko:2018eod,Tawfik:2014eba}. This could be understood in the light of the themal nature of an arbitrary small part of the highly entangled fireball states. Following the Eigenstate Thermalization Hypothesis \cite{Muller:2017vnp,DAlessio:2016rwt}, the corresponding probability distribution of the projection of these states remains thermal. We follow the line that the thermal models reproduce well the particle yields and the thermodynamic properties of the hadronic matter including the chiral and freezeout temperatures. We compare our calculations with reliable lattice QCD simulations, an effective QCD-like approach, and available experimental results. 

The present script is organized as follows. In section \ref{sec:Apprch} approaches for deconfinement and freezeout boundaries in equilibrium thermal models are introduced. The results are discussed in section \ref{sec:Rslt}. Section \ref{sec:Cncls} is devoted to the conclusions and outlook.

\section{Equilibrium thermal models}
\label{sec:Apprch}

It was conjectured that the formation of the hadron resonances follows the bootstrap picture, i.e. the hadron resonances or the fireballs are composed of further resonances or fireballs, which in turn are consistent of lighter resonances or smaller fireballs and so on \cite{Fast:1963uql,Fast:1963dyc}. The thermodynamic quantities of such a system can be deduced from the partition function $Z(T, \mu, V)$ of an ideal gas. In a grand canonical ensemble, this reads \cite{Tawfik:2014eba,Karsch:2003vd,Karsch:2003zq,Redlich:2004gp,Tawfik:2004sw,Tawfik:2005qh}
\bea 
Z(T,V,\mu)=\mbox{Tr}\left[\exp\left(\frac{{\mu}N-H}{T}\right)\right], \label{eq:lnZ} 
\eea
where $H$ is Hamiltonian combining all relevant degrees of freedom and $N$ is the number of constituents of the statistical ensemble. Eq. (\ref{eq:lnZ}) can be expressed as a sum over all hadron resonances taken from recent particle data group (PDG) \cite{Tanabashi:2018oca} with masses up to $2.5~$GeV \cite{Beringer:1900zz},
\begin{equation} 
\ln  Z(T,V,\mu)=\sum_i{{\ln Z}_i(T,V,\mu)} =V \frac{g_i}{2{\pi}^2}\int^{\infty}_0{\pm p^2 dp {\ln} {\left[1\pm {\lambda}_i \exp\left(\frac{-{\varepsilon}_i(p)}{T} \right) \right]}}, \label{eq:ln2}
\end{equation}
where the pressure reads $T\partial \ln  Z(T,V,\mu)/\partial V$, $\pm$ stands for fermions and bosons, respectively. $\varepsilon_{i}=\left(p^{2}+m_{i}^{2}\right)^{1/2}$ is the dispersion relation and $\lambda_i$ is the fugacity factor of $i$-th particle \cite{Tawfik:2014eba},
\begin{equation} 
\lambda_{i} (T,\mu)=\exp\left(\frac{B_{i} \mu_{\mathtt{b}}+S_{i} \mu_{\mathtt{S}}+Q_{i} \mu_{\mathtt{Q}}}{T} \right), \label{eq:lmbd}
\end{equation}
where $B_{i} (\mu_{\mathtt{b}})$, $S_{i} (\mu_{\mathtt{S}})$, and $Q_{i} (\mu_{\mathtt{Q}})$ are baryon strangeness, and electric charge quantum numbers (their corresponding chemical potentials) of the $i$-th hadron, respectively. From phenomenological point of view, the baryon chemical potential $\mu_{\mathtt{b}}$ can be related to the nucleon-nucleon center-of-mass energy $\sqrt{s_{\mathtt{NN}}}$ \cite{Tawfik:2013bza}
\bea
\mu_{\mathtt{b}} &=& \frac{a}{1+b \sqrt{s_{\mathtt{NN}}}}, \label{eq:mue}
\eea
where
$a=1.245\pm0.049~$GeV and $b=0.244\pm0.028~$GeV$^{-1}$. The number and energy density, respectively, can be derived as 
\begin{eqnarray}
n_i(T,\mu)&=&\sum_{i}\frac{\partial\, {\ln Z}_i(T,V,\mu)}{\partial\, \mu_{i}}=\sum_{i}\frac{g_{i}}{2\pi^{2}}\int_{0}^{\infty}p^{2} dp \frac{1}{\exp\left[\frac{{\mu}_{i} - {\varepsilon}_{i}(p)}{T}\right] \pm 1}, \label{eq:n} \\
\rho_i(T,\mu)&=&\sum_{i}\frac{\partial\, {\ln Z}_i(T,V,\mu)}{\partial\, (1/T)}=\sum_{i}\frac{g_{i}}{2\pi^{2}}\int_{0}^{\infty}p^{2} dp \frac{-\varepsilon_i(p)\pm\mu_i}{\exp\left[\frac{{\mu}_{i} - {\varepsilon}_{i}(p)}{T}\right] \pm 1}. \label{eq:e}
\end{eqnarray}
Likewise, the entropy and other thermodynamic quantities can be derived straightforwardly.

Both temperature $T$ and the chemical potential $\mu=B_{i} \mu_{\mathtt{b}}+S_{i} \mu_{\mathtt{S}}+\cdots$ are related to each other and to $\sqrt{s_{\mathtt{NN}}}$ \cite{Tawfik:2014eba}. As an overall thermal equilibrium is assumed, $\mu_{\mathtt{S}}$ is taken as a dependent variable to be estimated due to the strangeness conservation, i.e. at given $T$ and $\mu_{\mathtt{b}}$, the value assigned to $\mu_{S}$ is the one assuring $\langle n_{\mathtt{S}}\rangle-\langle n_{\bar{\mathtt{S}}}\rangle=0$. Only then, $\mu_{\mathtt{S}}$ is combined with $T$ and $\mu_{\mathtt{b}}$ in determining the thermodynamic quantities, such as the particle number, energy, entropy, etc. The chemical potentials related to other quantum charges, such as the electric charge and the third-component isospin, etc. can also be determined as functions of $T$, $\mu_{\mathtt{b}}$, and $\mu_{\mathtt{S}}$ and each of them must fulfill the corresponding laws of conversation. 

This research intends to distinguish between deconfinement and freezeout boundaries in equilibrium thermal models. The latter is characterized by $T_{\mathtt{\chi}}$ and $\mu_{\mathtt{b}}$, which are conditioned to one of the universal freezeout conditions \cite{Tawfik:2016jzk}, such as constant entropy density normalized to $T_{\mathtt{\chi}}^3$ \cite{Tawfik:2005qn,Tawfik:2004ss}, constant higher-order moments of the particle multiplicity \cite{Tawfik:2013dba,Tawfik:2012si}, constant trace anomaly \cite{Tawfik:2013eua} or an analogy of the Hawking-Unruh radiation \cite{Tawfik:2015fda}. The experimental estimation for $T_{\mathtt{\chi}}$ and $\mu_{\mathtt{b}}$, as shown in Fig. \ref{Fig:1} proceeds through statistical fits for various particle ratios calculated in statistical thermal models. The former, the deconfinement transition, is conditioned to line-of-constant-physics, such as constant energy density, $\rho$ \cite{Tawfik:2004vv}. The inclusion of the strange quarks seems to affect the critical temperatures, as these come up with extra hadron resonances and their thermodynamic contributions, where the mass of strange quarks is of the order of the critical temperature.

\section{Results}
\label{sec:Rslt}

\begin{figure}[thb] 
\begin{center}
\includegraphics[width=12.cm,angle=0]{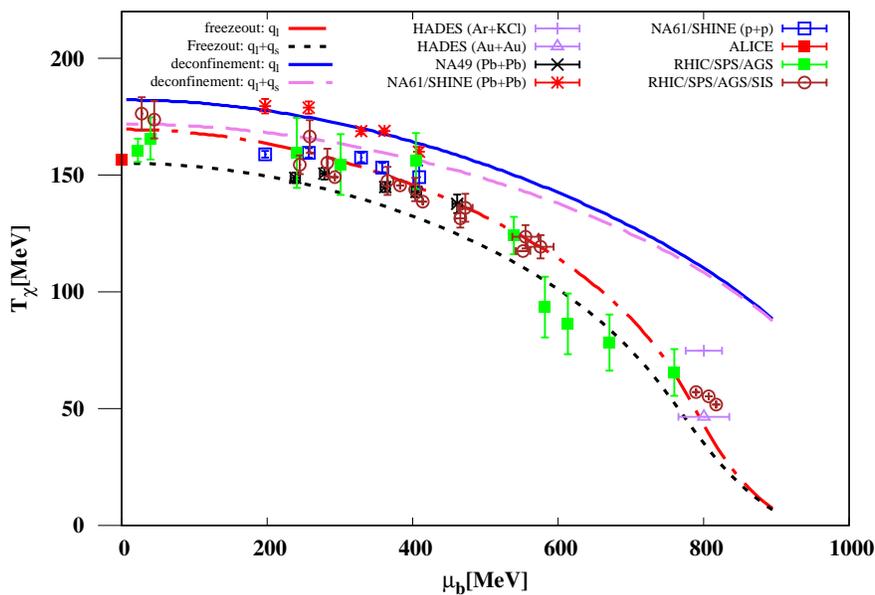} 
\caption{\footnotesize
The freezout and deconfinement parameters $T_{\mathtt{\chi}}$ and $\mu_{\mathtt{b}}$ as deduced from different experimental results (symbols with error bars) are depicted and confronted with the thermal model calculations for the chemical freezeout (dot-dashed and dashed curves) and the deconfinement parameters (solid and long-dashed curves) with and without strange quarks. \label{Fig:1}}
\end{center}
\end{figure}

Figure \ref{Fig:1} depicts the freezout and deconfinement parameters $T_{\mathtt{\chi}}$ and $\mu_{\mathtt{b}}$ as determined from the measurements gained from different experiments (symbols with error bars); HADES (Ar+KCal) \cite{agakishiev2011dielectron}, NA61/SHINE (Au+Au, p+p) \cite{Begun:2016pdy,Lorenz:2014eja,Abgrall:2013qoa,Agakishiev:2014nim,Agakishiev:2015ysr,Anticic:2009ie,anticic2009energy}, NA49/SHINE (Pb+Pb) \cite{vovchenko2017chemical,Alt:2004kq,Kraus:2004uk}, and  ALICE \cite{Floris:2014pta,andronic2018decoding,abelev2013k,alexandre2014multi,abelev2015k,alice20153,alice2017production,abelev2012pion} as well as analysis at RHIC/SPS/AGS \cite{Lorenz:2014eja} and RHIC/SPS/AGS/SIS energies \cite{cleymans2006comparison,adler2002centrality,adler2001multiplicity,adams2006identified,Velkovska:2001xz,Adams:2005dq,oeschler1999hadrons} are combined with each others and compared with the thermal model calculations. The latter take into account both freezeout (dot-dashed and dashed curves) and deconfinement boundaries (solid and long-dashed curves) with and without strange quarks. 

With the experimental results, we mean the parameters obtained when measured particle yields and/or ratios are fitted to calculations based on statisical thermal models, in which the parameters $T_{\mathtt{\chi}}$ and $\mu_{\mathtt{b}}$ are taken as independent variables, where the baryon-chemical potential $\mu_{\mathtt{b}}$, for instance, can directly be fixed at a given center-of-mass energy $\sqrt{s_{\mathtt{NN}}}$, Eq. (\ref{eq:mue}). For the freezeout parameters $T_{\mathtt{\chi}}$, $\mu_{\mathtt{b}}$, $\mu_{\mathtt{S}}$, etc. the thermodynamic quantities which fulfill one of the freezeout conditions reviewed in refs. \cite{Tawfik:2016jzk,Tawfik:2004sw}, such as constant entropy density normalized to temperature cubed.  Conditions for deconfinement phase transitions have been discussed in refs. \cite{Tawfik:2004vv,Tawfik:2004sw}, line-of-constant-physics, such as constant energy density with varying $\mu_{\mathtt{b}}$, $\mu_{\mathtt{S}}$, $\sqrt{s_{\mathtt{NN}}}$, etc. Details of the various approaches (curves) become in order now. With freezeout $q_l$ and freezeout $q_l+q_s$ we mean QCD phase boundaries as determined under freezeout conditions, where in the HRG model  is configured to have only hadrons whose constituents are light quarks ($q_l$) and to have only hadrons with light and strange quarks $q_l+q_s$. Under these conditions, it is likely to characterize the chemical freezeout $T-\mu_{\mathtt{b}}$-plane. Furthermore, with deconfine $q_l$ and deconfined $q_l+q_s$, we mean the line-of-constant-physics, which is defined by constant energy density, for instance. Such a line is mapped out in the HRG model in which only hadrons with light quarks ($q_l$) and only hadrons with light and strange quarks $q_l+q_s$ are included.
 
It is obvious that both sets of freezeout parameters seem identical, for instance, at low $\mu_{\mathtt{b}}$ or high $\sqrt{s_{\mathtt{NN}}}$, where the slight difference could be tolerated. At large $\mu_{\mathtt{b}}$ or low $\sqrt{s_{\mathtt{NN}}}$, the difference between the temperatures of freezeout and deconfinement becomes larger. Such a difference would be understood based on the assumption that the chemical freezeout takes place very late after the phase of hadronization. The latter is QCD confinement transition. Its order as simulated in recent lattice QCD is a likely crossover, i.e. there a wide range of temperatures within which QGP hadronizes or hadrons go through QGP. The time span becomes longer with the increase in $\mu_{\mathtt{b}}$ or the decrease in $\sqrt{s_{\mathtt{NN}}}$. The conjecture of the existence of a mixed phase is probably another possibility. In this phase, both types of degrees of freedom, hadrons and QGP, live together until the system goes through deconfinement to colored QGP or finally entirely freezes out to uncorrelated colorless hadrons.  

The co-existence of different QCD phases was discussed in litrature, for instance \cite{Yukalov:2013yj,Bugaev:2015vxa}. The mixed QCD phases can be formed in macroscopic, mesoscopic, and microscopic mixture. As shown in Fig. \ref{Fig:1}, these mixed phases start being produced at $\sqrt{s_{\mathtt{NN}}}$ ranging between $\sim 5$ and $\sim 12~$GeV, i.e. $\mu_{\mathtt{b}}\simeq 320$ to $\simeq 560~$MeV.

For the freezeout parameters, it is apparent that the agreement between the thermal model calculations and the experimental results is very convincing. This covers the entire $\mu_{\mathtt{b}}$-range and can - among other evidence - be interpreted based on the fact that the freezeout stage is the latest along the temporal evolution of the high-energy collision, where the number of produced particles is entirely fixed. The time elapsed from the stage of the chemical freezeout to the time of detection is likely shorter than the time from any other QCD processes, such as hadronization, chiral	symmetry breaking, etc. and therefore it is apparently the most accurate one.

In the present calculations, full quantum statistics \cite{Karsch:2003vd, Karsch:2003zq, Redlich:2004gp, Tawfik:2004sw, Tawfik:2005qh} and hadron resonances with masses up to $2.5~$GeV \cite{Beringer:1900zz} are taken into account. The strangeness degrees of freedom play an important role, especially at low $\mu_{\mathtt{b}}$ or high $\sqrt{s_{\mathtt{NN}}}$.  

For the sake of completeness, we have also checked the same calculations but with the inclusion of a large number of possible missing states \cite{ManLo:2016pgd, Noronha-Hostler:2016ghw}. We found that the thermodynamic quantities, especially the ones to which the present script is limited, show sensitivity to these missing states \cite{Capstick:1986bm}. They are entering our calculations in the same manner as done for the PDG hadrons and resonances. 

The missing states are resonances predicted, theoretically, but not yet confirmed, experimentally. Their quantum numbers and physical characteristics are theoretically well known \cite{Bazavov:2014xya}. Basically, they are conjectured to greatly contribute to the fluctuations and the correlations, i.e. higher derivatives of the partition function, estimated in recent lattice QCD simulations \cite{Bazavov:2014xya}. These are the occasions where their contributions becomes unavoidable \cite{ManLo:2016pgd}. Another reason for adding the missing states is that they come up with additional degrees of freedom and considerable decay channels even to the hadrons and resonances which are subject of this present study.

\begin{figure}[thb] 
\begin{center}
\includegraphics[width=12.cm,angle=0]{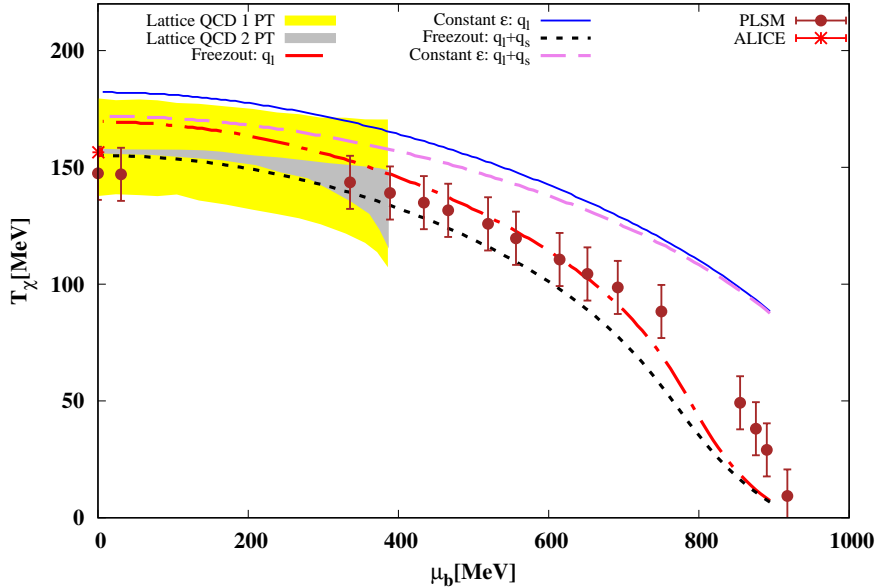} 
\caption{\footnotesize
The same as in Fig. \ref{Fig:1} but here a comparison between lattice QCD simulations \cite{Bellwied:2015rza,Bazavov:2018mes} and Polyakov linear-sigma model \cite{Tawfik:2014uka,Tawfik:2014gga,AbdelAalDiab:2018hrx,Tawfik:2015pqa,Tawfik:2016gye,Tawfik:2016edq,Tawfik:2019rdd} with the thermal model. \label{Fig:2} }
\end{center}
\end{figure}

For $T_{\mathtt{\chi}}$ and $\mu_{\mathtt{b}}$, a comprehensive comparison between the thermal model calculations (curves) and the results deduced from the lattice QCD simulations (bands)  \cite{Bellwied:2015rza,Bazavov:2018mes} and the Polyakov linear-sigma model (symbols with error bars) \cite{Tawfik:2014uka} is presented in Fig. \ref{Fig:2}. Within their statistical and systematic certainities, there is an excellent agreement between the lattice QCD simulations (bands) \cite{Bellwied:2015rza,Bazavov:2018mes} and the Polyakov linear-sigma model (symbols) \cite{Tawfik:2014uka}. The reason why the lattice QCD simulations are limited to $\mu_{\mathtt{q}}/T_{\mathtt{\chi}}\leq 1$ is the so-called sign problem and the difficulties which arise because of the importance of sampling becomes no longer possible. There are various attempts to anticipate this limitation; continuation from imaginary chemical potential, reweighting methods, applying complex Langevin dynamics, and Taylor expansions in the quark chemical potential $\mu_{\mathtt{q}}$ \cite{Philipsen:2010gj}. 

We also find an excellent agreement between the thermal model calculations for the chemical freezeout parameters and the predictions deduced from the Polyakov linear-sigma model, especially at $\mu_{\mathtt{b}}\gtrsim300~$MeV. At lower $\mu_{\mathtt{b}}$, the thermal model calculations seem to slightly overestimate $T_{\mathtt{\chi}}$.

This observed agreement would be taken as an evidence supporting the conclusion that the first-principle calculations likely result in  $T_{\mathtt{\chi}}$ - $\mu_{\mathtt{b}}$ plane similar to that of the Polyakov linear-sigma model, especially at large $\mu_{\mathtt{b}}$, where the first-principle calculations are no longer applicable.

It is in order now to highlight a few details of the linear-sigma model, which is much simpler than QCD, but based on QCD symmetries as well \cite{Tawfik:2014uka,Tawfik:2014gga,AbdelAalDiab:2018hrx,Tawfik:2015pqa,Tawfik:2016gye,Tawfik:2016edq,Tawfik:2019rdd}. Originally, it was intended to describe the pion-nucleon interactions and the chiral degrees of freedom. A spinless scalar field $\sigma_a$ and triplet pseudoscalar fields $\pi_a$ are introduced in theory of quantized fields to the linear-sigma model, which is a low-energy effective model, in which the generators $T_a=\lambda_a/2$ with Gell-Mann matrices $\lambda_a$ and the real classical field forming an ${\cal O}(4)$ vector, $\vec{\Phi}=T_a(\vec{\sigma}_a, i\vec{\pi}_a)$ are included. The chiral symmetry is explicitly broken by $3\times 3$ matrix field $H=T_a  h_a$, where $h_a$ are the external fields. Accordingly, under $SU(2)_L\times SU(2)_R$ chiral transformation, such as $\Phi\rightarrow L^+\Phi R$, the $\sigma_a$ sigma fields acquire finite vacuum expectation values, which in turn break $SU(2)_L\times SU(2)_R$ down to $SU(2)_{L+R}$. These transformations produce massive sigma particle and nearly massless Goldstone bosons, the pions. Therefore, the constituent quarks gain masses, as well; $m_q=g f_{\pi}$, where $g$ is coupling and $f_{\pi}$ is the pion decay constant. Also, the fermions can be introduced either as nucleons or as quarks. The $\sigma$ fields under chiral transformations exhibit the same behaviour as that of the quark condensates and thus, $\sigma$ can be taken as order parameters for the QCD chiral phase transition. 

With the incorporation of the Polyakov-loop potential so that the Lagrangian of the PLSM reads $\mathcal{L} = \mathcal{L}_{\bar{\psi}\psi}+\mathcal{L}_m-\mathbf{\mathcal{U}}(\phi, \phi^0, T)$, where the first term stands for Lagrangian density of fermions with $N_c$ color degrees-of-freedom, the second term gives the contributions of the mesonic fields, and finally the third term represents the Polyakov-loops potential incorporating the gluonic degrees-of-freedom and the dynamics of the quark-gluon interactions, i.e. deconfinement is also incorporated in this chiral model.

The questions which arise now are why PLSM reproduces well the non-perturbative lattice QCD simulations and why the PLSM agrees well with the thermal model calculations, especially for the freezeout boundary? The first question can be directly answered. PLSM incorporates both chiral and deconfinement QCD symmetries. On the other hand, it seems that both types of transitions are nearly coincident, especially at vanishing or small baryon chemical potentials. Within this region, both calculations are in excellent agreement with each others. At high temperatures, both chiral symmetry restoration and deconfinement transition produce almost free quarks and gluons, e.g. QGP. The reliability of the chiral effective model, PLSM, seems crucial, especially where lattice field theory is unavailable or the experimental results are not accessible yet. 

The second question about the reasons why PLSM agrees well with the freezeout parameters deduced from the thermal model calculations can be answered as follows. First, at $\mu_{\mathtt{b}}\gtrsim300~$MeV, where lattice field theory likely suffers from the sign problem, it seems that both chiral and deconfinement boundaries become more and more distinguishable. It might be obvious that the critical temperature of the chiral phase transition would be smaller than that of the deconfinement transition, which in turn differs from the freezeout temperatures. Within these two limits, which should be subject of further studies, a temperature region is created, in which a phase of mixed hadron-QGP likely takes place. Last but not least, the $T_{\mathtt{\chi}}$ - $\mu_{\mathtt{b}}$ plane of the Polyakov linear-sigma model \cite{Tawfik:2014uka} was determined under the condition of constant entropy density normalized to $T^3$ \cite{Tawfik:2016jzk,Tawfik:2005qn,Tawfik:2004ss}, i.e. likely manifesting the freezeout boundary. A future phenomenological study should be conducted in order to find out whether the condition of line-of-constant-physics gives results in agreement with the $T_{\mathtt{\chi}} - \mu_{\mathtt{b}}$ plane for deconfinement.

Now it is in order to summarize some details about the two lattice QCD calculations (yellow and grey bands). \cite{Bellwied:2015rza,Bazavov:2018mes}. In ref. \cite{Bellwied:2015rza}, the crossover boundary separating the hadron gas from the quark gluon plasma phases at small baryon chemical potentials was calculated using a four times stout smeared staggered fermion action with dynamical up, down, strange and charm quarks ($2+1+1$). This means that the two light quarks are degenerate. The masses of light quarks and that of the strange quark mass are tuned such that the physical pion and Kaon mass over the pion decay constant are reproduced for every lattice spacing. For the gauge action, the tree-level Symanzik improvement was used. In order to overcome the sign problem, imaginary values of the chemical potential have been considered, which are then translated through analytic continuation to the real values of the baryon-chemical potentials. The continuum extrapolation is based on lattices with $10$, $12$, and $16$ temporal dimensions. The curvature of $x=T_c(\mu_b)/T_c(\mu_b=0)$ was estimated according to anzaetze $1+ax$, $1+ax+bx^2$, $(1+ax)/(1+bx))$ and $(1+ax+bx)^{-1}$, where $a$ and $b$ are fit parameters. The corresponding critical temperature at $\mu_b=0$ was estimated as $157~$MeV, where vanishing strange density was assured.

In ref. \cite{Bazavov:2018mes}, the critical temperatures of chiral crossovers at vanishing and finite values of baryon ($b$), strangeness ($s$), electric charge ($Q$), and isospin ($I$) chemical potentials obtained in the continuum limit from lattice QCD calculations carried out for $2+1$ highly improved staggered quarks (HISQ) and the tree-level improved Symanzik gauge action with two degenerate up and down dynamical quarks and a dynamical strange quark, with physical quark masses corresponding to physical pion and Kaon masses are presented. The temporal extents are varied from $N_{\tau}=6$, $8$, $12$, and $16$, going towards progressively finer lattice spacing. The critical temperatures have been parameterized as $T_c(\mu_x)= T_c(\mu_x=0)\left[1 -\kappa_2^x(\mu_x/T_c(\mu_x))^2-\kappa_4x(\mu_x/T_c(\mu_x=0))^4 \right]$, where $\kappa_2^x$ and $\kappa_4^x$ are determined from Taylor expansions of chiral observables in $\mu_x$. The corresponding critical temperature at $\mu_b=0$ was estimated as $156.5\pm1.5~$MeV. At the chemical freeze-out of relativistic heavy-ion collisions and under thermal conditions, such as, $\mu_s(T,\mu_b)$ and $\mu_Q(T,\mu_b)$ determined from strangeness neutrality and isospin imbalance, $\kappa_2^b=0.012\pm 0.004$ and $\kappa_4^b=0.000 \pm 0.0004$.

\section{Conclusions and outlook}
\label{sec:Cncls}

Among the various phases which take place in the strongly interacting matter under extreme conditions, we focused on the deconfinement and chemical freezeout boundaries. The authors  compared results on $T_{\mathtt{\chi}}$ and $\mu_{\mathtt{b}}$ deduced from various heavy-ion experiments with recent lattice simulations, effective QCD-like Polyakov linear-sigma model, and equilibrium thermal model. Along the entire freezeout boundary, we conclude that an excellent agreement between the thermal model calculations and the experiments is found. Also, the estimations deduced from the Polyakov linear-sigma model excellently agree with the thermal model calculations. It should be noted that at low baryon density or high energies, both deconfinement and chemical freezeout boundaries are likely coincident. Accordingly, we can also conclude that  the lattice calculations for the deconfinement transition agree well with the Polyakov linear-sigma model, where in both approaches QCD symmetries are included.  At large baryon density or low energies, the two boundaries become distinguishable and probably form a phase in which hadrons and quark-gluon plasma likely coexist.
 
Based on the fact that Polyakov linear-sigma model agrees well with the lattice QCD simulations, at least within $\mu_{\mathtt{b}}$-range of reliable simulations, a future phenomenological study should be conducted on Polyakov linear-sigma model to find out whether the condition of line-of-constant-physics gives results in agreement with the $T_{\mathtt{\chi}} - \mu_{\mathtt{b}}$ plane for deconfinement. Furthermore, it intends to characterize the phase of mixed hadron-QGP and its possible predictions at the future facilities FAIR and NICA as well as its astrophysical consequences. 

\bibliographystyle{aip} 
\bibliography{ThmlMdlsDcnfFrz1}

\end{document}